\documentclass[a4paper,11pt]{article}
\pdfoutput=1

\usepackage{jcappub} 

\usepackage[T1]{fontenc}
\usepackage{multicol}
\usepackage[super]{nth}
\usepackage{subcaption}
\usepackage[font=small]{caption}
\newcommand{\fnl}{f_{\rm NL}}
\title{\boldmath Non-Gaussianity in inflationary scenarios for primordial black holes}

\author{Matthew W. Davies, Pedro Carrilho, David J. Mulryne}

\affiliation{Astronomy Unit, Queen Mary University of London,
Mile End Road, London, E1 4NS, United Kingdom}

\emailAdd{m.w.davies@qmul.ac.uk, p.gregoriocarrilho@qmul.ac.uk, d.mulryne@qmul.ac.uk}

\abstract{
Working in an idealised framework in which a series of phases of evolution defined by the second slow-roll parameter $\eta$ are matched together, we calculate the reduced bispectrum, $\fnl$, for models of inflation with a large peak in their primordial power spectra. We find $\fnl$ is typically approximately constant over scales at which the peak is located, and provide an analytic approximation for this value. This allows us to 
identify 
the conditions under which $\fnl$ is large enough to have a significant impact on 
the resulting production of primordial black holes (PBHs) and scalar induced gravitational waves (SIGWs). Together with analytic formulae for the gradient of the rise and fall in the power spectrum, this provides a toolkit for designing 
or quickly analysing inflationary models that produce PBHs and SIGWs.}

\begin{document}
\maketitle
\flushbottom

\section{Introduction}
Models of inflation capable of producing a large peak in the primordial power spectrum at small scales are the subject of much active research. At large scales, the power spectrum is tightly constrained by observations of the Cosmic Microwave Background (CMB) to be approximately scale-invariant, with a slight red tilt, and with an amplitude $\mathcal{A}_{\mathcal{R}}\sim\mathcal{O}(10^{-9})$ \cite{P2016,P20162}. These constraints do not apply, however, to the power at much smaller scales.

One potential consequence of an enhanced power spectrum is the formation of primordial black holes (PBHs). These form from the gravitational collapse of perturbations at enhanced scales upon horizon re-entry \cite{Carr}. Recently, there has been a renewed interest in PBHs, especially owing to their potential as a dark matter candidate \cite{Ivanov,Carr_2016,Carr_2017,Inomata_2017,Green_2021,Villanueva_Domingo_2021}. PBHs have also been suggested as a source for gravitational wave (GW) events witnessed by LIGO and Virgo collaborations \cite{Bird_2016,Sasaki_2016,Takhistov_2021,De_Luca_2021,Abbott_2021} and as contributors to the stochastic GW background which may have been detected by NANOGrav \cite{De_Luca_20212,Vaskonen_2021}.

If PBHs are to be produced in the numbers necessary to be responsible for these effects, then the power spectrum at small scales likely needs to be enhanced by $\sim {\cal O }(10^7)$  relative to CMB measurements of $\mathcal{A}_{\mathcal{R}}$ \cite{Zaballa_2007,Josan_2009,Motohashi_2017,Cole_2018,Germani_2019}, the precise enhancement required depending on the finer details  of both the collapse and the shape of the power spectrum. For power spectra with a large peak at short scales, scalar induced gravitational waves (SIGWs) will also be produced at second order in perturbation theory \cite{Ananda_2007,Baumann_2007,Saito_2009,Alabidi_2012,Nakama_2017,Garc_a_Bellido_2017,Kohri_2018,Di_2018,Clesse2018detecting,Bartolo_2019,Domenech:2021ztg}.

As well as the necessary growth in the power spectrum, it is well understood that non-Gaussianity of the scalar perturbations can have important effects on both PBH production \cite{Bullock_1997,2006hidalgo,hidalgo2007effect,Saito_2008,Byrnes_2012,Young_2013,Bugaev_2013,Young_2015,Young_2016,Franciolini_2018,Luca_2019,Passaglia_2019,atal2019role,taoso2021nongaussianities,Oz2021,rezazadeh2021nongaussianity} and the spectrum of SIGWs \cite{Cai_2019,Unal_2019,Yuan2020gravitational,Atal_2021,adshead2021nongaussianity,Ragavendra_2021,Domenech:2021ztg}. Although in reality the full probability distribution of perturbations is needed to accurately determine the production of PBHs, the reduced bipsectrum, $\fnl$, can be used as a guide for when non-Gaussianity becomes important to the calculation. An $\fnl$ of $\mathcal{O}(1)$ will significantly affect both the abundance of PBHs, and their mass distribution \cite{Byrnes_2012,Young_2013,Young_2016, Franciolini_2018, taoso2021nongaussianities}. In the case of SIGWs, if non-Gaussianity is local, then $\fnl$ together with the power spectrum is sufficient to fully determine the spectrum of GWs. In this context, for example, an $\fnl \sim \mathcal{O}(1) $ to $\mathcal{O}(10)$ will have a significant effect when  $\mathcal{A}_{\mathcal{R}} \sim 10^{-3}$ \cite{Cai_2019,adshead2021nongaussianity}.
Since PBHs and SIGWs form at enhanced scales, the most relevant value of $\fnl$ is its value around the peak of the power spectrum. 

There are many models for generating enhancements of the primordial power spectrum that have already been explored in the literature (see Refs. \cite{2020Moto,Tasinato_2021,taoso2021nongaussianities,inomata2021primordial,rezazadeh2021nongaussianity} for recent examples). In general, these are based on specific single-field inflationary potentials. A framework more amenable to analytic studies, however, was introduced by Byrnes et al. in Ref. \cite{Byrnes_2019}. Instead of working from a pre-determined potential,  inflation is modelled as a series of distinct "phases" each characterised by a constant value of the second slow-roll parameter $\eta$; transitions between phases occur instantly \cite{2015Moto,2020Moto}. This approach can be used to approximate a wide variety of models, and we refer to models in this framework as "idealised".

At present, works studying idealised models have focused on the power spectrum \cite{Byrnes_2019,Carrilho_2019,_zsoy_2020,Tasinato_2021,inomata2021primordial}. Our primary aim in this paper, therefore, is to extend this approach to compute the bispectrum as well, focusing in particular on the value of $\fnl$ at scales around the peak of the power spectrum. We find that $\fnl$ takes a similar form for all idealised models with a large peak in their power spectrum, and find a simple expression for the value of $\fnl$ at scales around this peak. We also generalise the idealised formalism further to allow $\eta$ to have a linear dependence on conformal time, using this to study the effect of relaxing the assumption of instantaneous transitions.

This paper is structured as follows: in Section \ref{sec:1} we present the general idealised framework for generating models of inflation with peaked primordial power spectra. In Section \ref{sec:2} we derive the form of the bispectrum using standard cosmological perturbation theory and the in-in formalism, $\fnl$ is also defined. In Section \ref{sec:3} we present plots of $\fnl$ in a selection of idealised models and derive a simple expression for its value around peak scales. In Section \ref{sec:4} we consider models with more realistic transitions which occur linearly rather than instantly in conformal time and discuss the effect of this on $\fnl$. Finally we summarise our results and discuss their implications for PBH formation and the spectrum of SIGWs.
\label{sec:intro}
\section{Models with Peaked Power Spectra}
\label{sec:1}
\subsection{Inflationary Action}
In this paper we consider single-field, minimally-coupled models of inflation defined by the action
\begin{equation} \label{Action}
S=\frac{1}{2} \int d^4 x \, \sqrt{-g} \left[R-(\nabla \phi)^2-2V(\phi)\right] \, ,
\end{equation}
where $g$ is the determinant of the metric, $R$ is the Ricci scalar, $\phi$ is the inflaton field and $V$ is its self-interaction potential. We define the  first and second slow-roll parameters as $\epsilon=-\frac{\dot{H}}{H^2}$ and $\eta=\frac{\dot{\epsilon}}{H \epsilon}$ respectively. A dot indicates a derivative with respect to cosmic time and $H$ is the Hubble parameter. It is assumed in all of our models that $\epsilon \ll 1$, but that $\eta$ can be much larger (at least of $\mathcal{O}(1)$, or higher). Using cosmological perturbation theory the action \eqref{Action} can be expanded in terms of the gauge-invariant comoving curvature perturbation $\mathcal{R}$ about a flat Friedmann-Robertson-Walker background. The power spectrum and bispectrum of  $\mathcal{R}$ is what we wish to compute.

\subsection{Mode Functions}
After expanding the action to second order in $\mathcal{R}$ we can obtain an equation of motion for the Fourier modes $\mathcal{R}_k$. Defining $z^2\equiv 2 a^2 \epsilon$, where $a$ is the scale factor and $v_k\equiv z \mathcal{R}_{k}$, the auxiliary field $v_k$ satisfies the Mukhanov-Sasaki equation \cite{Sasaki:1986hm,MUKHANOV1992203}
\begin{equation}
    v''_k+\left(k^2-\frac{z''}{z}\right)v_k=0 \, ,
\end{equation}
where a dash indicates a derivative with respect to conformal time $\tau$. This equation can be rewritten in terms of the quantity
\begin{equation}
    \nu^2=\frac{9}{4}+\frac{3}{2}\eta+\frac{1}{4}\eta^2+\frac{\dot{\eta}}{2H}+\mathcal{O}(\epsilon)
\end{equation}
as
\begin{equation}
v''_k+\left(k^2-\frac{\nu^2 -\frac{1}{4}}{\tau^2}\right)v_k=0 \, ,
\label{eq:M-S}
\end{equation}
where we have used the de-Sitter space approximation $aH=-1/\tau$. For constant $\eta$, 
the general solution to this equation is written in terms of Hankel functions of the first and second kind as
\begin{equation} \label{MF}
\mathcal{R}_k=\frac{-H\tau}{\sqrt{2\epsilon(\tau)}}\left[C_1 \sqrt{-\tau} H^{(1)}_{\nu}(-k \tau)+C_2 \sqrt{-\tau}H^{(2)}_{\nu}(-k\tau)\right] \, .
\end{equation}
Note also that $\epsilon$ depends on conformal time as $\epsilon \propto (-\tau)^{-\eta}$ for constant $\eta$.

\subsection{Idealised Approach}
\label{approach}
Following the approach of Byrnes et al. in Ref. \cite{Byrnes_2019}, we can now use this solution to construct models of inflation with peaked power spectra in our idealised framework. We model inflation as a succession of phases defined by a constant $\eta$ and assume that transitions between these phases occur instantaneously. This means that the solutions for the mode functions in each phase are given by Eq. \eqref{MF}, with constants $C_1$ and $C_2$ taking different values in each phase. In the first phase the constants are fixed by assuming the Bunch-Davies initial conditions at some initial time, and in subsequent phases by requiring continuity of ${\cal R}$ and its first derivative at each transition time, according to the Israel junction conditions \cite{Israel:1966rt,Deruelle_1995}.

\subsection{Linear Transitions}
\label{linear}
If we want to consider more realistic transitions that are not instantaneous, we must allow for $\eta$, and hence $\nu$, to be time dependent. 
If $\eta$ varies linearly 
in conformal time, an analytical solution to Eq.~\eqref{eq:M-S} exists, allowing us to continue 
with a fully analytic 
approach. The solution is given in 
terms of Whittaker functions as
\begin{equation} \label{MFL}
\mathcal{R}_k=\frac{(-H\tau)}{\sqrt{2\epsilon(\tau)}}[C_1 M_{p, q}(r)+C_2 W_{p,q}(r)] \, ,
\end{equation}
where the values of $p,q$, and $r$ depend on the details of the transition. If we take a general linear transition between two values of $\eta$, $\eta_1$ and $\eta_2$, then we have
\begin{equation} \label{Lin}
    \eta(\tau)=\eta_1 + (\eta_2 - \eta_1)\frac{\tau-\tau_1}{\tau_2-\tau_1} \, ,
\end{equation}
where the transition between $\eta_1$ and $\eta_2$ takes place between the conformal times $\tau_1$ and $\tau_2$, $\tau_1<\tau_2$. In this case the values of $p,q,r$ are:
\begin{align}
    p&=-\frac{(\eta_1-\eta_2)(-\tau_2(2+\eta_1)+\tau_1(2+\eta_2))}{4(\tau_1-\tau_2)^2\sqrt{-k^2+\frac{(\eta_1-\eta_2)^2}{4(\tau_1-\tau_2)^2}}} \, , \\
    q&=\frac{1}{2}\sqrt{\frac{(\tau_2(3+\eta_1)-\tau_1(3+\eta_2))^2}{(\tau_1-\tau_2)^2}}\, , \\
    r&=2\left(\sqrt{-k^2+\frac{(\eta_1-\eta_2)^2}{4(\tau_1-\tau_2)^2}}\right)\tau\,.
\end{align}
$C_1$ and $C_2$ are once again fixed by continuity. 

In this paper, in addition to instantaneous transitions, we consider models where $\eta$ varies linearly in conformal time during a transition with non-zero duration. We refer to these as "linear transition" models.

\subsection{Power Spectrum}
The power spectrum $P_{\mathcal{R}}(k)$ for a given inflationary model can be calculated from the mode functions. It is defined by its relation to the two-point correlator
\begin{equation}
   \langle \mathcal{R}_{\mathbf{k}}\mathcal{R}_{\mathbf{k}'}\rangle=(2\pi)^3\delta(\mathbf{k}+\mathbf{k}')P_\mathcal{R}(k) \, .
\end{equation}
The dimensionless power spectrum $\mathcal{P}_{\mathcal{R}}(k)$ is then defined in terms of the power spectrum as
\begin{equation}
    \mathcal{P}_{\mathcal{R}}(k)=\frac{k^3}{2\pi^2}P_{\mathcal{R}}(k) \, .
\end{equation}
In this paper, ``power spectrum'' will always refer to the dimensionless power spectrum, unless otherwise specified. This can be expressed neatly in terms of the mode functions as
\begin{equation}
    \mathcal{P}_\mathcal{R}(k)=\frac{k^3}{2 \pi^2}|\mathcal{R}_k|^2 \, .
\end{equation}

\subsection{Scenarios with a Large Peak}
\label{sec:scenarios}

As mentioned, we are interested in models 
of inflation which produce a large peak in the power spectrum.
Sharp peaks can be generated in a variety of ways, but 
in general they require the first slow-roll 
parameter to decay rapidly (see for example the discussion in Ref. \cite{Carrilho_2019}). The simplest case is therefore to begin in a phase of slow-roll, with negligible $\eta$, and transition rapidly to 
a phase with $\eta\ll 0$. When $\eta<-3$,  
rather than being frozen, modes grow on superhorizon scales. This not only affects modes which exit the horizon during the the large negative $\eta$ phases, but also modes which exited during some period before it began. The number of modes affected
is determined by the length of the negative $\eta$ phase. As described in Ref.~\cite{Carrilho_2019} it is these modes that gain a large blue spectral index, and lead to 
a steep power spectrum. Modes that exit
during slow-roll initially freeze, but 
can then grow again later 
during a large negative $\eta$ phase
and gain a $\mathcal{P}_\mathcal{R}(k) \propto k^4$ dependence~\cite{Byrnes_2019}. This is almost the steepest sustained $k$-dependence the power-spectrum can achieve 
(so long as there is only one large negative $\eta$ 
phase --- see Appendix~\ref{App1} for a full discussion); a marginally steeper growth of 
$\mathcal{P}_\mathcal{R}(k) \propto k^5(\log k)^2$ is
possible if an $\eta = -1$ phase precedes the large negative $\eta$ phase~\cite{Carrilho_2019}. In general, modes that exit the horizon before a 
large $\eta$ phase which they are affected by gain a $k$-dependence of the form~\cite{Carrilho_2019}
\begin{equation}
\mathcal{P}_\mathcal{R}(k) \propto k^{5 -| \eta +1 |}\,,
\label{eq:tiltA}
\end{equation}
where $\eta$ in this 
formula is the $\eta$-value of the phase in which modes exit. 
The scale dependence is as described by Eq.~\eqref{eq:tiltA} so 
long as the growing phase lasts a sufficiently long time for the initially 
sub-leading term of the asymptotic expansion of the mode functions 
to become dominant~\cite{Carrilho_2019}. This 
compares to the standard expression
\begin{equation}
\mathcal{P}_\mathcal{R}(k) \propto k^{3 -| \eta +3 |}\,,
\label{eq:tilt}
\end{equation} 
for modes whose spectral index
is not affected by future phases of evolution, which follows from the leading term in an 
asymptotic expansion of the mode functions.
We review and refine the discussion of Ref.~\cite{Carrilho_2019} in Appendix~\ref{App1}. We now highlight a selection of scenarios in which a large power spectrum peak can be generated.

\paragraph{The standard scenario}
The simplest way to get a large peak in the 
power spectrum follows from the discussion above. 
We begin in a phase of slow-roll $\eta\sim0$ inflation in order to be consistent with CMB observations of approximate scale-invariance at large scales. During 
this period of slow-roll inflation 
the inflaton rolls down its self-interaction potential. This is followed by a period with a substantial 
decrease in the kinetic energy of the inflaton. In terms of $\eta$ values this corresponds to a phase with a large, negative value of $\eta$. The
simplest case of  $\eta=-6$ is referred to as ``ultra-slow-roll'' (USR) and is one of the examples most-studied in the literature. This phase is achieved by introducing an extremely flat region in the inflaton's potential. Modes which exit during USR have a scale invariant spectrum, and so, in order bring the power spectrum down again after the peak, a final phase with a large positive $\eta$-value is required. This corresponds to a steep fall off 
in the inflationary potential and to the inflaton gaining kinetic energy again. Modes which exit 
during this phase freeze 
after horizon crossing in the usual way. 
It isn't necessary for the power spectrum's amplitude to be reduced back to CMB levels, but a model must end in a phase of $\eta>-3$ otherwise super-horizon modes will continue to evolve. 
This phase could be followed by other 
phases, such as a further slow-roll phase, but 
as long as there are no more 
phases with $\eta<-3$ the modes
which exited in previous phases are unaffected. 

We can 
also change the value of $\eta$ in the large negative $\eta$ phase so that we get something other than USR. Values with  $\eta<-6$ correspond 
to the inflaton slowing down more quickly 
than it would on a flat potential, and 
require the potential to start to rise.
This option is appealing 
because modes which exit  
during this phase have a red spectral index, and 
the spectrum begins to 
decay from its peak due to these modes alone (i.e.
without necessarily needing modes to exit in a  
subsequent large positive $\eta$ phase). A further phase of $\eta > 0$ is still required, however, so that the modes freeze.

In summary, in the "standard scenario" we have three phases with $\eta\sim 0$, $\eta \leq-3$, and $\eta > 0$. 
Modes which exit towards 
the end of the slow-roll phase ultimately gain a $k^4$ dependence, as predicted by Eq.~\eqref{eq:tiltA}; modes which exit during the $\eta<-3$ phase gain a scale dependence  as predicted by Eq.~\eqref{eq:tilt};
and modes which exit in the large 
positive $\eta$ phase are red tilted also as predicted by Eq.~\eqref{eq:tilt}. 
The overall appearance is a  sharp rise followed by a sharp fall, although possibly with an extended peak in between depending on the value of the large negative $\eta$. In the simplest case of USR, there is an extended, flat peak. 

\paragraph{Variants on the standard scenario}

It is, of course, possible to generalise the 
standard scenario. We can consider 
adding a further phase after the slow-roll phase but before the large negative 
$\eta$ phase. Modes exiting during these 
phases will be affected by the large negative $\eta$ phase and lead to a different growth than the $k^4$ behaviour. 
If, for example, an $\eta =-1$ phase is added, the slightly steeper growth of $\mathcal{P}_\mathcal{R}(k) \propto k^5(\log k)^2$ is found as mentioned above.

\paragraph{Inflaton Falls} 

One particular variant of the standard scenario 
is noteworthy, and we refer to it as the "inflaton falls" model after the study 
of Inomata et al. in Ref. \cite{inomata2021primordial}.
In this case an initial phase of slow-roll inflation is followed immediately 
by a phase with a large positive value of $\eta$, after which there is a large negative $\eta$ stage, in that order, before the model returns back to slow-roll (or indeed 
any other $\eta>-3$ value). This corresponds to the inflaton "falling" off its slow-roll plateau and 
its kinetic energy initially increasing, before
reaching a sufficiently flat (or growing) region 
of the potential in which the inflaton decelerates again. If the large negative $\eta$ phase is large enough to affect not only 
the modes which exit immediately before 
it during the large positive $\eta$ phase, but also 
some of the modes that exit at the end of the 
slow-roll phase, the result is that the 
modes which exit during slow-roll gain a $k^4$ dependence as usual, and the modes which exit during the large $\eta$ phase gain 
a dependence according to Eq.~\eqref{eq:tiltA}. With a large positive $\eta$, this implies a strongly red, decaying spectrum (but with a different tilt than if these modes simply freeze and the tilt was determined by Eq.~\eqref{eq:tilt}).

In Ref. \cite{inomata2021primordial}
the discussion implies that the peak in the power spectrum is produced by a rapid increase in the kinetic energy of the inflaton thanks to a sharp dip in its potential, but this hides the origin of the peak somewhat. Really it is modes that exit during 
slow-roll which ultimately account for the steep growth to a peak, and the modes which exit during the phase in which the kinetic energy increases which 
account for a steep decay. 
As in the standard scenario the large negative $\eta$ phase is required to generate the
steep rise (by changing the tilt of the slow-roll modes). 

\paragraph{Repeating Model} In Ref. \cite{Carrilho_2019} it is argued that 
only a marginally steeper sustained growth 
than $k^4$ could possibly be generated during canonical inflation. 
This statement remains true if there is only one large negative $\eta$ phase, but in later work Tasinato \cite{Tasinato_2021} 
showed that a faster growth can be achieved by successive phases of SR  
and USR.
We attempt to give a full explanation of how 
this works in Appendix~\ref{App1}. 
This model begins in a phase of SR, before switching twice between phases of USR and SR. It can be generalised to have other values of $\eta$ in these 
alternating phases, and will also require a large positive $\eta$-value added in one of the ways described in the simpler scenarios above 
in order for the spectrum to decay.
\\

We will now compute the bispectra for each of the  kinds of model just introduced. 
The remarkable thing we find is that, 
despite the numerous possibilities for 
generating a large peak in 
the power-spectrum and the differences in the 
precise form of the spectrum between them, in all of them the reduced bispectrum, $\fnl$, looks remarkably similar and possesses the same structure. 
As discussed, of particular interest is the non-Gaussianity at scales around the peak of these power spectra.
We will show that the level of non-Gaussianity around the peak, in all of these models, can be approximately calculated from the same extremely simple formula.

\section{Bispectrum}
\label{sec:2}
\subsection{Third Order Action}
To compute the bispectrum we need to expand the action \eqref{Action} to third order in the curvature perturbation \cite{Maldacena_2003, Namjoo_2013}
\begin{equation} \label{COA}
\begin{aligned}
S_3={}&\int dt \, d^3x \, \bigg[a^3 \epsilon^2 \mathcal{R}\dot{\mathcal{R}}^2+a\epsilon^2\mathcal{R}(\partial \mathcal{R})^2-2a \epsilon \dot{\mathcal{R}}(\partial \mathcal{R})(\partial \chi)\\
&+\frac{a^3 \epsilon}{2}\dot{\eta}\mathcal{R}^2\dot{\mathcal{R}}+\frac{\epsilon}{2a}(\partial\mathcal{R})(\partial\chi)\partial^2 \chi+\frac{\epsilon}{4a}(\partial^2 \mathcal{R})(\partial \chi)^2\\
&+2f(\mathcal{R})\frac{\delta L}{\delta \mathcal{R}}\bigg\rvert_1 \bigg] \, ,
\end{aligned}
\end{equation}
where
\begin{equation*}
\partial^2 \chi=a^2 \epsilon \dot{\mathcal{R}}\, , \,  \frac{\delta L}{\delta \mathcal{R}}\bigg\rvert_1=a(\partial^2 \dot{\chi}+H\partial^2 \chi - \epsilon \partial^2 \mathcal{R}) \, ,
\end{equation*}
and
\begin{equation} \label{FOR}
\begin{aligned}
f(\mathcal{R})={}&\frac{\eta}{4}\mathcal{R}^2+\frac{1}{H}\mathcal{R}\dot{\mathcal{R}}\\&+\frac{1}{4 a^2 H^2}[-(\partial \mathcal{R})(\partial \mathcal{R})+\partial^{-2}(\partial_i \partial_j(\partial_i \mathcal{R} \partial_j \mathcal{R}))]\\&+\frac{1}{2a^2 H}[(\partial \mathcal{R})(\partial \chi)-\partial^{-2}(\partial_i \partial_j(\partial_i \mathcal{R}\partial_j\chi))] \, .
\end{aligned}
\end{equation}
The last term in the cubic action is usually removed by making a field redefinition \cite{Maldacena_2003} of the form
\begin{equation} \label{FR}
    \mathcal{R}\rightarrow \mathcal{R}_n + f(\mathcal{R}_n) \, .
\end{equation}
Once the final term has been removed, the remaining terms in Eq. \eqref{COA} are at least of $\mathcal{O}(\epsilon^2)$ or higher with the exception of fourth term,  which is proportional to $\epsilon \dot{\eta}$. In our models $\eta$ varies rapidly between phases of inflation (either instantaneously, 
or via rapid linear transitions as discussed in \S\ref{linear}), meaning that contributions from 
this term dominate over all others. 

Calculation of the bispectrum 
can now proceed using the redefined  
quantity. In the 
final answer, however, we 
must remember to convert back to 
the original physical curvature perturbation, and hence $f(\mathcal{R}_n)$ will also contribute to the bispectrum. In the models we consider in this work we will only need to retain the first term in $f(\mathcal{R}_n)$. This is because evaluation 
of the final bispectrum will always take place at late times, once the scales being considered are super-horizon, and during a phase in which $\mathcal{R}$ is conserved on super-horizon scales (a phase where $\eta>-3$). This means the third term can be neglected due to spatial derivatives, while the second and fourth terms can be neglected as both are proportional to $\dot{\mathcal{R}_n}$.  After these considerations, we end up with just two contributions to the bispectrum: one coming from the only surviving term in the cubic order action
\begin{equation}
\label{cubic}
S_3 \approx \int dt d^3 x \frac{a^3\epsilon}{2} \dot{\eta}\mathcal{R}_n^2 \dot{\mathcal{R}_n}
\end{equation}
and the other coming from the first term in the field redefinition
\begin{equation}
\label{FRT}
f(\mathcal{R}_n) \approx \frac{\eta}{4}\mathcal{R}_n^2 \, .
\end{equation}
\subsection{Form of the Bispectrum}
The bispectrum  $B(k_1, k_2, k_3)$ can be expressed in terms of the three-point correlator of the mode functions
\begin{equation} \label{bispec}
    \langle \mathcal{R}_{\mathbf{k}_1}\mathcal{R}_{\mathbf{k}_2} \mathcal{R}_{\mathbf{k}_3}\rangle=(2\pi)^3 B(k_1,k_2,k_3)\delta(\mathbf{k}_1+\mathbf{k}_2+\mathbf{k}_3) \, .
\end{equation}
This can be computed by using the 
in-in formalism to 
calculate the three point function 
of the redefined field and then 
accounting for the field redefinition using Wick's theorem \cite{Maldacena_2003}. The contribution from Eq. \eqref{cubic} (dropping the subscript $n$ for simplicity) is given by the integral \cite{Chen} (see also \cite{2013Moto})
\begin{equation} \label{cubicbi}
B_{\rm int}(k_1,k_2,k_3)= -2 \, \Im \,\mathcal{R}_{k_1}(\tau_0)\mathcal{R}_{k_2}(\tau_0)\mathcal{R}_{k_3}(\tau_0) \int_{-\infty}^{\tau_0} d\tau \, a^2 \epsilon \eta^\prime \left[ \mathcal{R}^\ast_{k_1}(\tau)\mathcal{R}^\ast_{k_2}(\tau)\mathcal{R}^{\ast \prime}_{k_3}(\tau) +\textrm{perm.}\right]\,,
\end{equation}
where $\mathcal{R}_{k_i}(\tau)$ are 
the mode functions to be calculated in the idealised approach of \S\ref{approach}. 

Since we are interested in the bispectrum evaluated at late times, $\tau_0$ will be some conformal time long after the transitions have taken place and all modes of interest have left 
the horizon. 
In the case of 
instantaneous transitions 
the integral Eq. \eqref{cubicbi} becomes a sum \cite{Chen}
\begin{equation} \label{cubicbiSum}
B_{\rm int}(k_1,k_2,k_3)= -2 \,  \Im \,\mathcal{R}_{k_1}(\tau_0)\mathcal{R}_{k_2}(\tau_0)\mathcal{R}_{k_3}(\tau_0) \sum_i  a(\tau_i)^2 \epsilon(\tau_i) \Delta\eta_i \left[ \mathcal{R}^\ast_{k_1}(\tau_i)\mathcal{R}^\ast_{k_2}(\tau)\mathcal{R}^{\ast \prime}_{k_3}(\tau_i) +\textrm{perm.}\right]\, ,
\end{equation}
where $\tau_i$ labels the time 
of the $i$th transition, and $\Delta \eta_i$ is the change in $\eta$ at this 
transition.

Next we account for the field redefinition. For a field redefinition of the form: $\mathcal{R}\rightarrow \mathcal{R}_n+q \mathcal{R}_n^2$, Wick's Theorem tells us that
\begin{equation}
\langle\mathcal{R}(x)\mathcal{R}(y)\mathcal{R}(z)\rangle=\langle\mathcal{R}_n(x)\mathcal{R}_n(y)\mathcal{R}_n(z)\rangle+q[\langle\mathcal{R}(x)\mathcal{R}(y)\rangle\langle\mathcal{R}(x)\mathcal{R}(z)\rangle+\textrm{perms.}]\,.
\end{equation}
In our case the first term here is the contribution from the cubic order action of our redefined field given above, and $q =\eta/4$.
Returning to Fourier space, the contribution of the field redefinition to the bispectrum can be written in terms of mode functions $\mathcal{R}_k$ as
\begin{equation} \label{redef1}
    B_{\rm fr}(k_1,k_2,k_3) = (2\pi)^3\frac{\eta}{4}\left[|\mathcal{R}_{k_1}(\tau_0)|^2|\mathcal{R}_{k_3}(\tau_0)|^2+\textrm{perms.}\right]\,.
\end{equation}
The sum of Eq. \eqref{cubicbi} and  Eq. \eqref{redef1} gives the full bispectrum for any model of the form we're considering.

Finally, we define the reduced bispectrum $\fnl(k_1,k_2,k_3)$, which is useful a measure of the relative size of non-Gaussianity
\begin{equation} \label{fnl}
    \frac{6}{5}\fnl(k_1,k_2,k_3)=\frac{B(k_1,k_2,k_3)}{P(k_1)P(k_2)+P(k_1)P(k_3)+P(k_2)P(k_3)} \, .
\end{equation}

\section{Non-Gaussianity in Idealised Models of Inflation}
\label{sec:3}
\begin{figure}
\centering
\begin{subfigure}{0.45\textwidth}
    \centering
    \includegraphics[width=\textwidth]{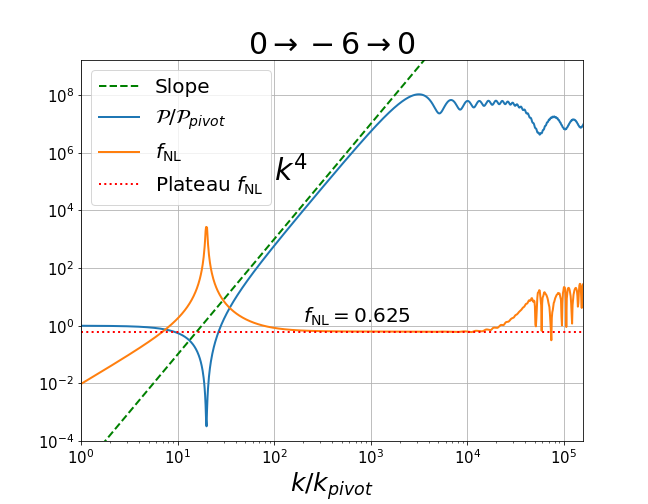}
\end{subfigure}
\begin{subfigure}{0.45\textwidth}
    \centering
    \includegraphics[width=\textwidth]{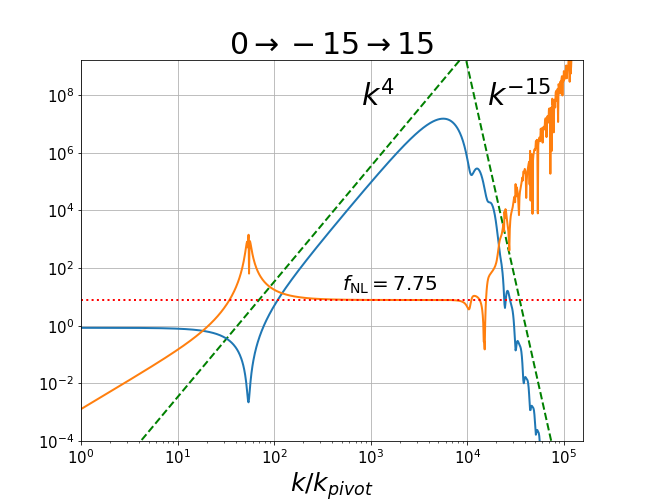}
\end{subfigure}
\begin{subfigure}{0.45\textwidth}
    \centering
    \includegraphics[width=\textwidth]{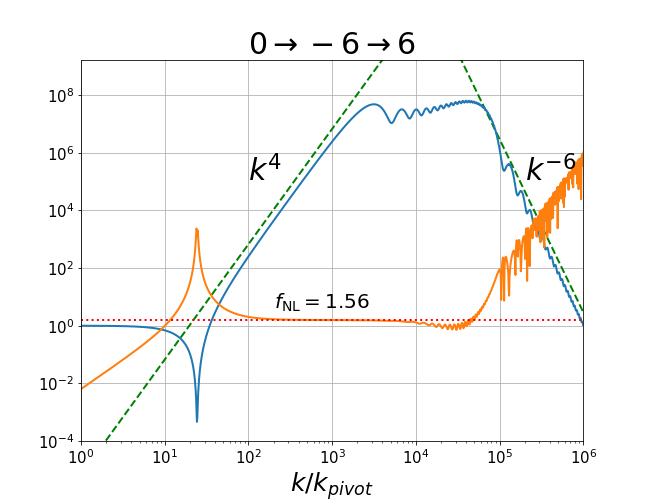}
\end{subfigure}
\begin{subfigure}{0.45\textwidth}
    \centering
    \includegraphics[width=\textwidth]{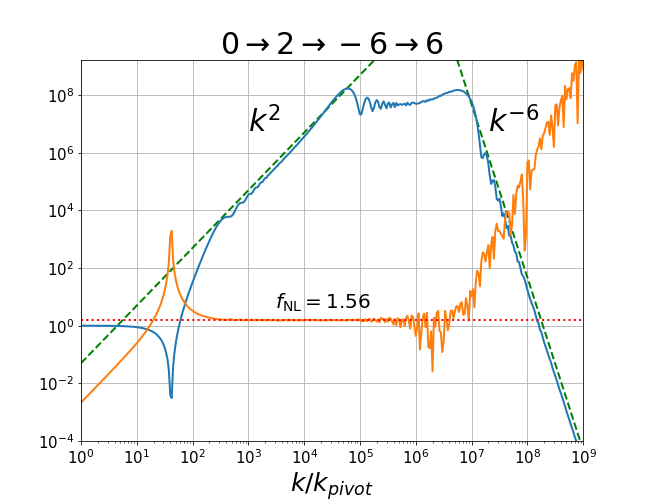}
\end{subfigure}
\begin{subfigure}{0.45\textwidth}
    \centering
    \includegraphics[width=\textwidth]{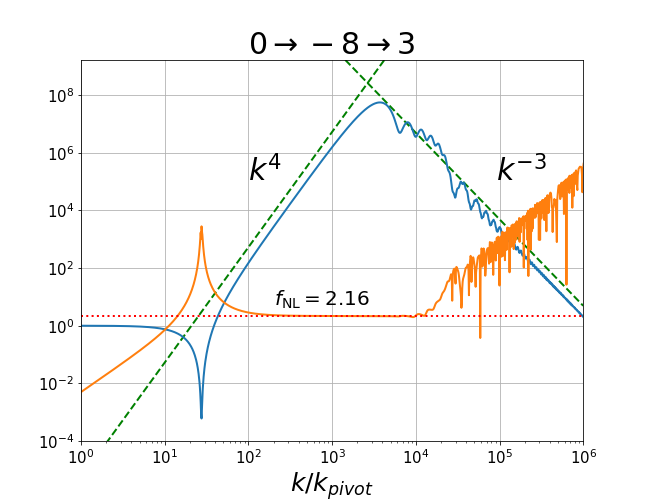}
\end{subfigure}
\begin{subfigure}{0.45\textwidth}
    \centering
    \includegraphics[width=\textwidth]{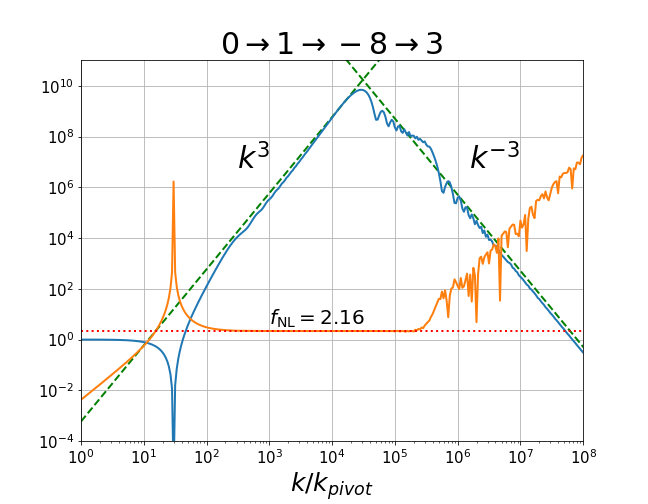}
\end{subfigure}
\caption[short]{Plots of the power spectrum and $\fnl$ (in the equilateral configuration) as a function of wavenumber $k$ for models of inflation in the standard scenario with instant transitions between three phases. Three additional models which are variants on the standard scenario are also shown. These include the commonly studied SR to USR to SR model, as well as two models with four phases which demonstrate how the scale-dependence of the rise in the power spectrum can be altered without affecting the value of $\fnl$ around the peak. The titles of the plots refer to the value of $\eta$ in the first, second and third (and possibly fourth) phase respectively. The power spectrum is plotted in blue, $\fnl$ in orange. Also displayed are the slopes of the rise and fall of the power spectrum as predicted by Eq. \eqref{eq:tiltA} or Eq. \eqref{eq:tilt} (dashed green), and the plateau value of $\fnl$ as predicted by Eq. \eqref{analytic} (dotted red). Note that the scale dependence of the rise in the power spectrum is given by Eq. \eqref{eq:tiltA}, whereas the scale dependence of the fall is given by Eq. \eqref{eq:tilt}. For a full discussion of scale dependence see \S\ref{sec:scenarios}.}
 \label{fig:instant}
\end{figure}
\begin{figure}
\centering
\begin{subfigure}{0.45\textwidth}
    \centering
    \includegraphics[width=\textwidth]{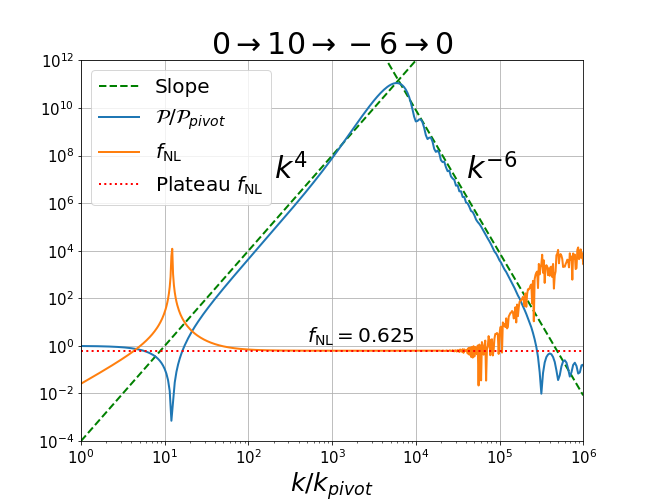}
\end{subfigure}
\begin{subfigure}{0.45\textwidth}
    \centering
    \includegraphics[width=\textwidth]{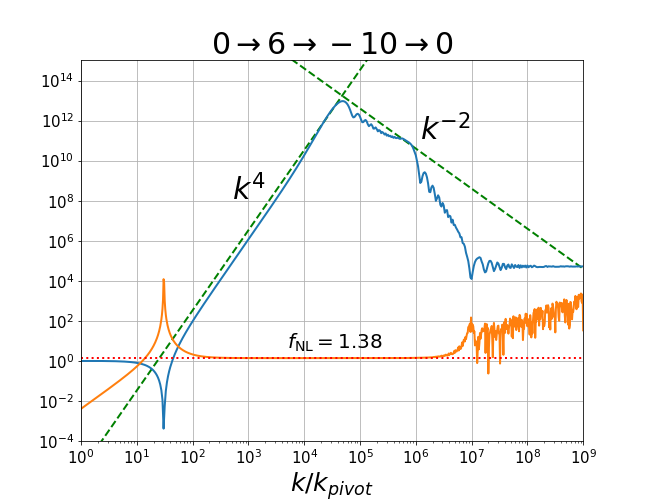}
\end{subfigure}
\begin{subfigure}{0.45\textwidth}
    \centering
    \includegraphics[width=\textwidth]{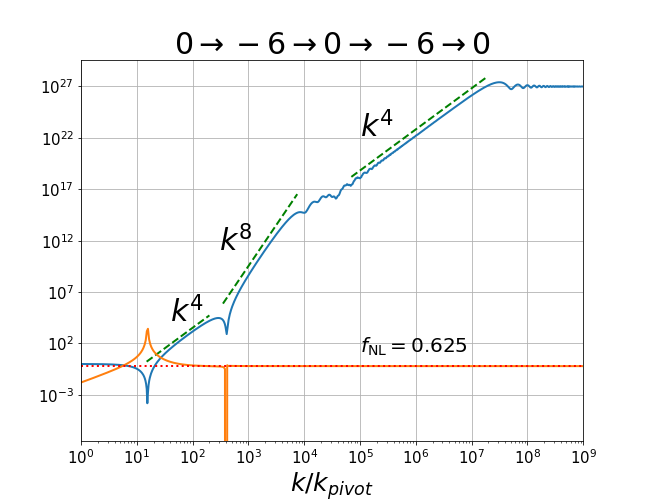}
\end{subfigure}
\caption[short]{Plots of the power spectrum and $\fnl$ (in the equilateral configuration) as a function of wavenumber $k$ for two "inflaton falls" models (top left and top right) and a repeating model (bottom) with instant transitions between the phases. The power spectrum is plotted in blue, $\fnl$ in orange. As in the previous figure, the slopes of the power spectrum and the plateau $\fnl$ value are also shown. In the case of an inflaton falls model, the scale dependence of the rise and the fall of the power spectrum are both given by Eq. \eqref{eq:tiltA}. The scale dependence of the rise in the repeating model is more subtle and discussed in Appendix \ref{App1}.}
 \label{fig:IF}
\end{figure}
Using the expressions given above we now calculate the value of $\fnl$ for a selection of idealised inflationary models, as described in \S\ref{sec:scenarios}. We consider first the case of 
instantaneous transitions between 
phases, and focus on 
the equilateral configuration\footnote{For clarity, we emphasise that we are plotting the amplitude of $\fnl$ (as defined in Eq. \eqref{fnl}) in the equilateral configuration. This is \textbf{not} the amplitude of the equilateral shape of $f_{\rm NL}^{\rm eq}$. If the shape is close to local (as we find it to be in the plateau region) then the value of $\fnl$ in the equilateral configuration coincides with $f_{\rm NL}^{\rm loc}$ \cite{planckcollaboration2019planck}} $(k_1=k_2=k_3=k)$. Remarkably, we 
find that in all models investigated 
that give a very large peak, the non-Gaussianity consistently exhibits the same structure. Moreover, at scales around the peak of the power spectrum, $\fnl$ is quasi-local (we observe this by investigating $\fnl$ away 
from the equilateral limit, finding only weak dependence on configurations for these scales). 

To present the results\footnote{In some plots the reader may be concerned by the size of the enhancement of the power spectrum, and also the possibility that the constraint $\fnl \lessapprox 1/P_{\cal{R}}$ may be violated leading to a strong-coupling problem. Please note that several of these examples are not intended to be fully realistic, but merely to illustrate clearly the behaviour of certain quantities and the utility of the approximate expressions we have developed. In realistic examples one would need to convert the power spectrum and bispectrum to an estimate for the production of PBHs (and SIGWs) ensuring overproduction does not occur. One must also respect the constraint that the power spectrum remain less than unity, and the strong coupling bound mentioned.}, in Fig. \ref{fig:instant} we plot both the power spectrum and the value of $\fnl$ for a selection of models which produce power spectrum peaks in the standard scenario and variations. In Fig. \ref{fig:IF} we do the same for an inflaton falls model, and a repeating model. In all of these cases the structure of $\fnl$ looks very similar; the features common to all these plots include: 
\begin{itemize}
    \item A large peak for modes exiting during the initial phase --- this corresponds to the dip in the power spectrum where the leading and sub-leading terms in the mode functions cancel each other out;
    \item A long, approximately flat region that begins for modes exiting during the initial phase and ends for modes exiting during the final phase. We refer to this as the "plateau" region, and the value of $\fnl$ within it is what we call the "plateau value". It extends over the scales at which the peak in the power spectrum occurs;
    \item A growing, highly oscillatory region at short scales.
\end{itemize}
Similar features, such as a scale-invariant $\fnl$ over peak scales, have also been noted in previous studies e.g. \cite{atal2019role}.

It is already known that the presence of sharp features in the inflaton's self-interaction potential  will lead to oscillations in $\fnl$ at very small scales \cite{Chen_2010}, in our models this is due to the instantaneous nature of the transitions between phases. These oscillations can clearly be seen at short scales in our plots in Fig. \ref{fig:instant}, for example. They begin to appear at scales $k \sim -1/\tau_n$ where $\tau_n$ is the conformal time at which the final transition between phases occurs. This means that they appear for modes exiting the the horizon during the final phase of a given model, which in our case will always be a phase with $\eta>-3$ where the power spectrum is decaying. Consequently, these oscillations will always appear at scales smaller than those at which the peak of the power spectrum is located. These are scales smaller than those we are interested in.

One can also see in Fig. \ref{fig:instant} that once the oscillations in $\fnl$ begin at short scales, the amplitude of these oscillations appears to grow without bound. This is a result of dealing with models in which the transitions between phases are instantaneous; such a growth is not a feature of realistic physical models where oscillations will be damped on short scales.

Of most interest is the value of $\fnl$ in the plateau region. Since the plateau extends over the scales at which the power spectrum reaches its peak, it is this value of $\fnl$ which is important for the formation of PBHs and SIGWs. It is natural, therefore, to ask what determines its value. We find that, as long as the large negative $\eta$ phase
lasts sufficiently long to form a peak of the size needed 
to produce PBHs, a plateau region in $\fnl$ always forms. Moreover, we see
that the value of $\fnl$ in this region is sensitive only to the large negative $\eta$-value, and the value of $\eta$ in the phase immediately following it. The plateau value is 
almost completely 
independent of the $\eta$-values of any previous phases, or their 
duration.

\subsection{Plateau Value}
An important feature of the 
plateau region of $\fnl$ is 
that it extends across modes which 
exited the horizon in different phases. For example, in standard scenarios it includes modes which 
exited during the large negative $\eta$ phase, and the phase immediately prior. 

To obtain an approximate 
expression for $\fnl$ 
in the plateau phase, therefore, we 
use this observation and 
study $k$-modes which exit the horizon during a phase of $\eta_1<-3$, and are matched at the transition to a phase of $\eta_2>-3$ evolution. Since ${\cal R}$ is ultimately 
conserved 
in this second phase, as long as this phases last sufficiently 
long, any further phases that 
also conserve ${\cal R}$ cannot 
affect the answer. 
We also assume that such modes are unaffected by transitions while deep inside the horizon 
and so 
are normalised to the Bunch-Davis vacuum at early times.
Under these assumptions, the answer for the two phase model 
gives us an approximate value for the $\fnl$ in the plateau of a model where $\eta_1$ is the $\eta$-value in its most recent negative $\eta$ phase, and $\eta_2$ the $\eta$-value of the phase immediately following it.

To find an analytic expression for $\fnl$ using 
the the two-phase model, the general mode functions \eqref{MF} for both phases are expanded in the superhorizon limit ($k \tau \rightarrow 0$). 
Using these expansions of the mode functions the power spectrum and bispectrum are computed as described 
above.
The bispectrum receives contributions from two terms: one from the single instantaneous transition in this 
model, and the other from the field redefinition term. It is clear that the redefinition term provides a scale-invariant contribution $f_{\rm NL,~redef}=\frac{5}{12} \eta_2$ to the non-Gaussianity \cite{atal2019role, taoso2021nongaussianities}, while 
the contribution from the transition is harder to estimate\footnote{In \cite{atal2019role}, for a more restricted class of models, it was shown that the contribution from the transition is negligible and that the only contribution to $\fnl$ is the scale-invariant one from the field redefinition.}.

We proceed by considering the full expression obtained from both contributions using Mathematica. The expression is rather long and unwieldy, but it can be simplified drastically by taking suitable limits. First, we expand the reduced bispectrum in terms of $\tau_0$, the time at which it is evaluated. Since we are interested in the non-Gaussianity at times long after the transitions have taken place, only the leading order $\mathcal{O}(\tau_0^0)$ term of this expansion is retained.

A barrier to further simplification is the presence of gamma functions in the resulting expression. This can be 
overcome by initially restricting to integer values of $\eta$ and dealing with 
odd and even values separately. Subsequently, the $k \rightarrow 0$ limit is applied to reveal a very neat expression. Finally, we verify that 
the result holds for non-integer $\eta$. The 
expression is given by 
\begin{equation} \label{analytic}
  \fnl =  \frac{5(9+4\eta_2 + \eta_1 (2+\eta_2))}{4(\eta_1-\eta_2)}\,.
\end{equation}
Recall that this estimate is for $\eta_1<-3$ and $\eta_2>-3$, and hence that it breaks down as $\eta_1 \to \eta_2$ where the expression exhibits a singularity.

A few further comments are in order. As discussed above if additional phases are present after the 
$\eta_2$ phase, they do not affect 
the power spectrum or bispectrum as long as they conserve ${\cal R}$. We note that in such cases
the scale invariant 
contribution to $\fnl$ from the field redefinition term alters in value, but this change is precisely accounted for by contributions 
from the additional transition terms. 

It only remains then to understand why this approximation works for modes 
which exit the horizon in the phase immediately 
before the large negative $\eta$ phase, and why all 
other transitions before this phase
do not contribute to the bispectrum.
We believe the former is true because the only dependence on the previous phases comes from the previous transition term in Eq.~\eqref{cubicbiSum}, which is negligible when compared to the final transition term as well as to the field redefinition terms, for scales that are affected by the rapid growth. The reason for this can be seen from the time dependence of the transition term in Eq.~\eqref{cubicbiSum}. The three copies of ${\cal R}$ are evaluated long after the transition and, have thus grown far beyond the amplitudes of the other copies of ${\cal R}$ evaluated at the transition time. However, when computing $f_{\rm NL}$, we divide by the final power spectrum, which effectively includes four copies of ${\cal R}$. This ratio is thus proportional to $1/{\cal R}$ meaning it decays as ${\cal R}$ grows. The same does not happen for the final transition term because all copies of ${\cal R}$ have similar sizes, since the phase after the transition conserves ${\cal R}$. It is also clear 
that earlier transitions provide even smaller contributions to the bispectrum, since the amplitude of ${\cal R}$ which enters Eq.~\eqref{cubicbiSum} is 
so much smaller before modes enter the $\eta <-3$ growing 
phase.

Finally, we note that the virtue of the expression in Eq. \eqref{analytic} is that it can be used to rapidly estimate the value of $\fnl$ around the peak of a power spectrum in a wide range of inflationary models. Below, we discuss the implications of this expression for $\fnl$ for a few different classes of models present in the literature.

\subsection{Standard Scenario}
\begin{figure}
\centering
\begin{subfigure}{.4\textwidth}
  \centering
  \includegraphics[width=\linewidth]{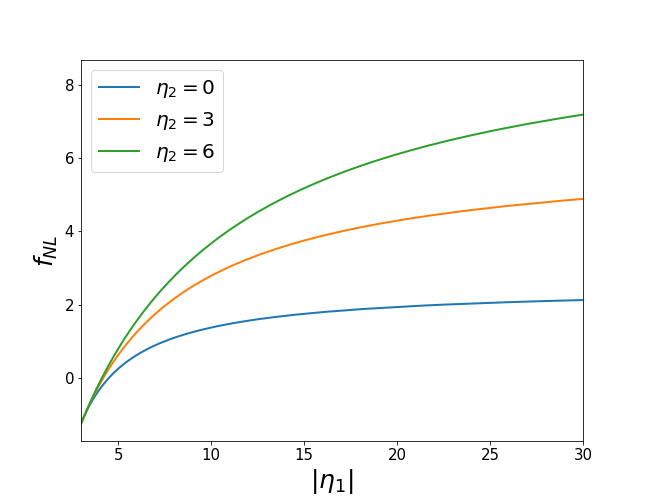}
\end{subfigure}
\begin{subfigure}{.4\textwidth}
  \centering
  \includegraphics[width=\linewidth]{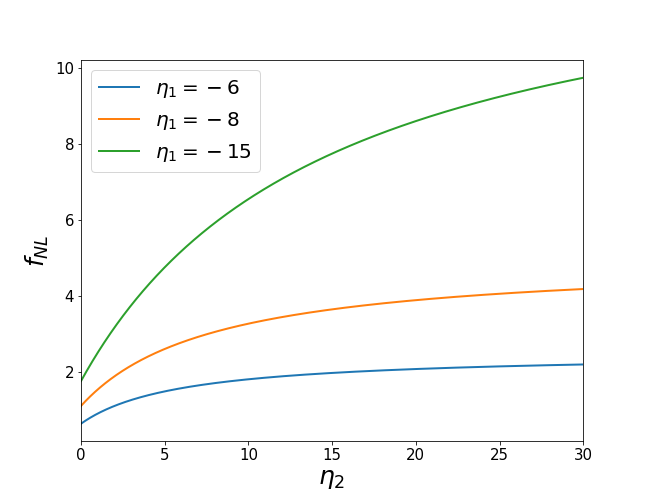}
\end{subfigure}
\begin{subfigure}{.4\textwidth}
  \centering
  \includegraphics[width=\linewidth]{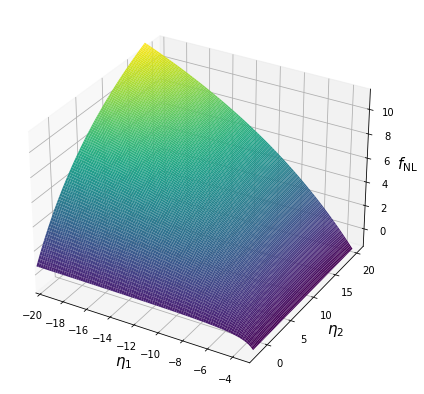}
\end{subfigure}
\begin{subfigure}{.4\textwidth}
  \centering
  \includegraphics[width=\linewidth]{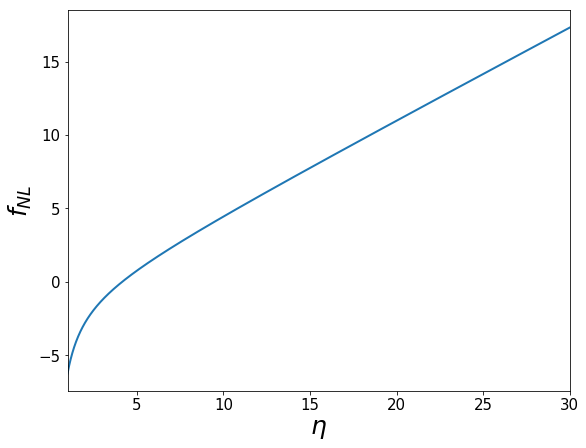}
\end{subfigure}
\caption{Dependency of $\fnl$ on $\eta$-values on the relevant phases of inflation. In general, $\eta_1$ is the most recent large negative $\eta$-value of the model and $\eta_2$ the $\eta$-value of the phase immediately following. In all cases, the value of $\fnl$ is larger when the magnitude of any $\eta$-values are increased. When one $\eta$ is kept fixed, as in a (top left) and b (top right), $\fnl$ tends to a certain value as the other $\eta$'s magnitude is increased. In c (bottom left) one can see a section of the 3D parameter space. In d (bottom right) we see the dependence of $\fnl$ on $\eta$ in a model of the form $0\rightarrow -\eta \rightarrow \eta$. When $\eta$ is increased $\fnl$ grows with no bound on its value.}
\label{fig:results}
\end{figure}

In Fig. \ref{fig:instant}, in addition
to the full $\fnl$, we plot the approximate plateau value
obtained by substituting the values of $\eta$ during the final two phases into Eq. \eqref{analytic}. Note that we have also verified that this value does not depend on the initial phase being SR (any other phase, such as $\eta = -1$ will lead to the same plateau value of $\fnl$), and that extra $\eta>-3$ phases can be glued to the end of the model also without affecting the plateau value. 

Next, in Fig. \ref{fig:results}, we use Eq. \eqref{analytic} to see what effect varying the values of $\eta$ in the final two phases will have on the plateau $\fnl$. If we fix one of these two values, and vary the other, then the plateau $\fnl$ grows with the magnitude of the varying $\eta$. The greater the magnitude of $\eta$, the greater the plateau $\fnl$, but only up to a limit. When one of the $\eta$ values is kept constant there is a maximum $\fnl$ that can be achieved. If we fix $\eta$ in the middle phase to be $\eta_m$, then the maximum non-Gaussianity that can be obtained is $-\frac{5(4+\eta_m)}{4}$. This is the value of $\fnl$ that would be obtained if the perturbations in the growing mode phase froze immediately at the end of this phase \cite{Namjoo_2013,Chen}. In other words, the limiting value of $\fnl$ in standard scenario models is the value of $\fnl$ if one only had the growing phase.

In addition, if we fix $\eta$ in the final phase to be $\eta_f$, then the maximum value of $\fnl$ is $\fnl=\frac{5(2+\eta_f)}{4}$. There is, however, no limit to the value of $\fnl$ if both $\eta$ values are allowed to vary. The value of $\fnl$ can be made arbitrarily large by increasing the magnitude of both.

\subsection{Inflaton Falls Models}
In Fig. \ref{fig:IF} (top left and top right), we plot the value of $\fnl$ for two inflaton falls models. In these cases the approximate plateau
value is still obtained by using Eq. \eqref{analytic}, except now substituting $\eta_2 =0 $ (since the growing phase is followed by slow-roll) and  whatever the value of $\eta$ is in the third phase of the model (the one with a large negative $\eta$ value) for $\eta_1$. We have verified that the plateau value of $\fnl$ does not depend at all on the large positive $\eta$ value, or on the fact that the model begins in a phase of SR inflation.

This means that, so long as the slow-roll phase immediately after the growing phase is retained, there is a limit on the $\fnl$ that can be achieved in an inflaton falls model. Since $\eta_2$ is fixed to 0 in Eq.~\eqref{analytic}, the maximum $\fnl$ is obtained in the limit as $\eta_1 \rightarrow -\infty$. In this limit, $\fnl=\frac{5}{2}$. If this slow roll phase is instead replaced by a different $\eta>-3$ phase then this value should be used for $\eta_2$ in Eq.~\eqref{analytic} and values of $\fnl$ larger than $\frac{5}{2}$ can, in principle, be obtained.

\subsection{Repeating Models}
In repeating models a faster than $k^4$ growth in the power spectrum is achieved thanks to the alternating SR and USR phases. However, once again, the non-Gaussianity has the same structure in repeating models as for the inflaton falls and standard scenario models. A long plateau in $\fnl$ extends over scales at which the peak in the power spectrum occurs, and the value of $\fnl$ here is given by substituting $\eta_1=-6$ and $\eta_2=0$ into Eq. \eqref{analytic}. This is the same non-Gaussianity that one would expect for a standard scenario model with a final transition from USR to SR.

Note that the value of $\fnl$ around the peak of such a model still only depends on the final large negative $\eta$-value and the $\eta$-value of the phase immediately following. A $0 \rightarrow -6 \rightarrow 0 \rightarrow -6 \rightarrow 0$ model will have the same $\fnl$ around its peak as a $0 \rightarrow -10 \rightarrow 0 \rightarrow -6 \rightarrow 0$, for example. The significance of the first negative $\eta$-value is for the scale dependence of the rise in the power spectrum, not for the value of $\fnl$. 

\section{Linear Transitions}

\label{sec:4}
\begin{figure}
    \centering
    \includegraphics[width=\textwidth]{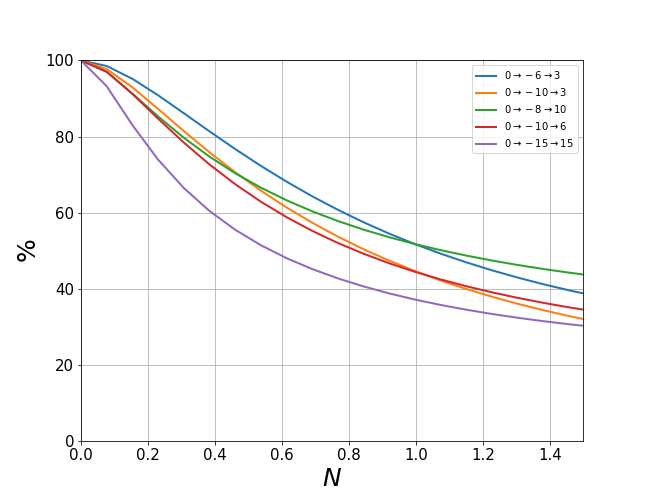}
    \caption{The dependence of the plateau value of $\fnl$ on the duration of the final transition in linear models. On the $y$-axis we plot the $\fnl$ value in the linear model as a percentage of its value in the instant case. The $x$-axis gives the number of $e$-folds the final transition lasted.}
    \label{fig:linear}
\end{figure}
Up until this point we have only been considering models of inflation with instant transitions between phases. While this simplified our analysis, these kinds of transition are unphysical approximations to realistic, continuous transitions. To investigate how our results are affected if we don't assume instantaneous transitions, we performed bispectrum calculations for models with transitions of finite duration between phases where the value of $\eta$ varied linearly in conformal time. This way the value of $\eta$ is continuous throughout the evolution.

We say that these models have "linear" transitions, since the value of $\eta$ evolves according to Eq. \eqref{Lin} during the transition. The setup and 
mode functions during the transitions 
were described in \S \ref{linear}, while the bispectrum calculation proceeds by integrating Eq.~\eqref{cubicbi} with 
the mode functions having a dependence on $\tau$. 

What we notice is that, although the reduced bispectrum in the linear case has a similar form to that in the instant case, the value of $\fnl$ in the plateau region is always reduced in the linear case. The reduced value of $\fnl$ depends only on the duration of the transition between the two phases whose $\eta$-values would be used in Eq. \eqref{analytic} to calculate $\fnl$ in the instant case. The details of any other linear transitions in the model do not affect $\fnl$. The longer the relevant transition lasts, the smaller the plateau value of $\fnl$ compared to its value in the instant case.

In Figure \ref{fig:linear} we plot the dependence of $\fnl$ on the duration of the transition for a selection of models in the standard scenario. In most cases studied $\fnl$ is reduced to around 50\% of its value for transitions lasting longer than an $e$-fold. In most cases $\fnl$ is reduced at most to 90\% of its value for transitions lasting less than a fifth of an $e$-fold. For models with larger $\eta$-values however, the decline in $\fnl$ is steeper. 

Our findings agree with earlier work that indicates that the value of $\fnl$ is very sensitive to the details of the transitions between phases \cite{Chen}. Slow, smooth transitions are liable to wipe out large amounts of non-Gaussianity. So long as the final transition is sufficiently short (much less than an $e$-fold), however, our formula gives a good estimate for $\fnl$ around the peak of the power spectrum.

\section{Conclusions}

In this work we have shown that the level of non-Gaussianity around the peak of a power spectrum, as measured by $\fnl$, can be predicted in general idealised models of inflation using a simple expression. This expression can be applied to estimate $\fnl$ in any model which has a correspondent in our idealised framework.

An $\fnl$ of $\mathcal{O}(1)$, or higher, can have a significant impact on PBH formation and the spectrum of SIGWs produced. In our framework there is, in principle, no limit to the magnitude of $\fnl$ that can be obtained. The larger the magnitude of the relevant $\eta$-values, the larger $\fnl$. $\mathcal{O}(1)$ values of $\fnl$ can be obtained for fairly reasonable $\eta$-values; a model with $\eta_1=-8$ and $\eta_2=10$ has $\fnl=3.26$, a model with $\eta_1=-30$ and $\eta_2=30$ has $\fnl=17.3$.

Our idealised models are clearly not fully realistic, since transitions between phases occur instantaneously. When we smooth out transitions by giving them a non-zero duration, we find invariably that $\fnl$ is reduced. If the transitions are long and smooth, this is liable to wipe out a significant fraction of $\fnl$, as predicted by our simple expression. So long as the transitions are short (much less than an $e$-fold), then our expression is still a good estimate for $\fnl$. 
It is worth asking whether short 
transitions between phases are realistic when constructing models directly from the inflationary potential. The answer requires investigation beyond the scope of this work, but it is clear that for some transitions this likely can be 
realised, while for others it would require much more fine tuning. For example transitioning rapidly into a slow-roll phase requires the field velocity 
to match precisely that of the 
of the slow-roll predicted value. More likely is that the field enters a regime of the potential that supports slow-roll, but with incorrect velocity, and there is a decay time-period over which the field velocity decays or grows to reach the slow-roll attractor. On the other hand, moving from to a phase of very large positive or 
negative $\eta$ can likely happen very rapidly. Such phases are defined by a large acceleration or deceleration of the field, and this only requires the potential to change suddenly such that it applies the required accelerating (or decelerating) force on the field. We defer more detailed investigation to future work, and conclude by reiterating we expect that our formula \eqref{eq:tiltA} for the tilt of the 
power spectrum (and the standard expression \eqref{eq:tilt}), together with our formula for $\fnl$ \eqref{analytic}, will help identify the important features of the spectra 
produced by models of inflation with a large peak, and aid in the designing of such models.

\appendix
\section{Detailed explanation of scaling of power spectrum}\label{App1}

In this appendix, we describe the scaling of the power spectrum induced by a fast growing stage.

We begin by re-writing Eq.~\eqref{MF} as
\begin{equation} \label{MF_gen}
\mathcal{R}_k^\eta=(-\tau)^{3/2+\eta/2}\left[A_\eta  H^{(1)}_{\nu}(-k \tau)+B_\eta H^{(2)}_{\nu}(-k\tau)\right]\,,
\end{equation}
which becomes, in the super-horizon limit $k\tau\rightarrow 0$
\begin{equation} \label{MF_exp}
\mathcal{R}_k^\eta=\tilde {B_\eta}k^{-3/2-\eta/2}\left(2+2\eta+(k\tau)^2\right)+\tilde {A_\eta}k^{3/2+\eta/2}(-\tau)^{3+\eta}\,,
\end{equation}
where the variables with tildes are combinations of $A_\eta$ and $B_\eta$ depending only on $\eta$. It is clear here that the constant mode is always present in the $\tilde{B_\eta}$ term, but time-dependent contributions can arise in both terms. In particular, for $-3<\eta<-1$ the leading decaying mode comes from the $\tilde {A_\eta}$ term, whereas for $\eta>-1$, the slowest decaying mode comes from the $\tilde {B_\eta}$ contribution. However, when a growing mode is present ($\eta<-3$), it only arises in the $\tilde {A_\eta}$, making it the crucial component for stages like USR. 

Matching a stage with constant leading behaviour (e.g. SR) to one with growing behaviour (e.g. USR) at $\tau=\tau_1$, one finds the following solution for the $\tilde {A_\eta}$ term of the second stage
\begin{equation} \label{A2_sol}
k^{3/2+\eta_2/2}\tilde{A}_{\eta_2}=k^2k^{-3/2-\eta_1/2}\left(\bar B_{\eta_1}+\bar A_{\eta_1}(k\tau_1)^{1+\eta_1}\right)\,,
\end{equation}
where the barred variables are proportional to the tilde variables. It can be easily checked that, at leading order in the limit $k\tau_1\rightarrow 0$, this expression always has the same scale dependence as the leading decaying term in the previous stage, as shown in Ref.~\cite{Carrilho_2019}. Squaring the expression in Eq.~\eqref{A2_sol} and multiplying by $k^3$ gives the scaling of the power spectrum $\mathcal P\sim k^\alpha$ as 
\begin{equation} 
\alpha=5-|\eta_1+1|\,.\label{scaling1}
\end{equation}

A hidden assumption used in deriving the result shown above is that the coefficients $A_{\eta_1}$ and $B_{\eta_1}$ are scale-independent. This is true for general values of $\eta$ in the case of Bunch-Davies initial conditions, but may not be correct if other transitions have occurred before the first stage considered here. This is particularly important if there are several stages with growing behaviour, as has been considered by Ref.~\cite{Tasinato_2021}, finding a scaling that does not obey Eq.~\eqref{scaling1}. We will now proceed to explain the origin of this effect.

After an USR-like stage it is often useful to match to an SR-like stage, in which the scale dependence of the growing term is naturally passed to the constant mode in the SR-like stage. What is crucial is the fact that the coefficient of the constant mode in the SR-like stage ($\tilde {B_\eta}$) is the same as the leading decaying term (for $\eta>-1$). And if that coefficient gains an additional scale dependence due to a previous USR-like stage, then so does the leading decaying term, contrary to what happens with Bunch-Davies initial conditions. For this reason, when there is a second USR-like stage following the second SR-like stage, the scaling of the power spectrum becomes modified. 

To demonstrate this in more detail, we perform these two additional matching calculations. We are assuming a scenario with three transitions in which the values of $\eta$ obey $\eta_1,\eta_3>-3$ and $\eta_2,\eta_4<-3$.
It is clear from Eq.~\eqref{A2_sol}, that one would need to know the scale dependence of both coefficients in the previous stage. Under the assumption that $\eta_1>-3$, we find
\begin{equation} \label{B2_sol}
\tilde{B}_{\eta_2}\approx \bar B_{\eta_1} k^{(\eta_2-\eta_1)/2}\,,
\end{equation}
where once more $\bar B_{\eta_1}$ is a place-holder variable that is proportional to $\tilde B_{\eta_1}$. It is not equal to the one shown in Eq.~\eqref{A2_sol}, but we are interested here only on how these coefficients depend on each other, not their specific expressions. After the second transition, the coefficients are
\begin{align} 
\tilde{A}_{\eta_3}\approx &\bar A_{\eta_2} k^{(\eta_2-\eta_3)/2}\,, \label{A3_sol}\\
\tilde{B}_{\eta_3}\approx &\bar A_{\eta_2} k^{3+(\eta_2+\eta_3)/2}+\bar B_{\eta_2} k^{(\eta_3-\eta_2)/2}\,.\label{B3_sol}
\end{align}
By multiplying by the scale dependence of the constant mode and neglecting the $\bar B_{\eta_2}$ term, we find
\begin{equation} \label{B3_sol_scaling}
k^{-3/2-\eta_3/2}\tilde{B}_{\eta_3}\approx \bar A_{\eta_2} k^{3/2+\eta_2/2}\,,
\end{equation}
where we can see that this mode inherits the same scale dependence of the previous growing mode, as expected (the right-hand side above is the same as the left-hand side of Eq.~\eqref{A2_sol}). To finalize, we put everything together into Eq.~\eqref{A2_sol}, since $\tilde A_{\eta_4}$ obeys the same equation. For $\eta_3>-1$, we find 
\begin{equation} \label{A4_sol1}
k^{3/2+\eta_4/2}\tilde{A}_{\eta_4}\approx k^2\left(k^{3/2+\eta_2/2}\bar A_{\eta_2}+\bar B_{\eta_2}k^{-3/2-\eta_2/2}\right)\,,
\end{equation}
where we can see that the contribution of the first growing stage is multiplied by an additional $k^2$, thus potentially increasing the steepness of the final power spectrum. Performing the full substitution and multiplying by an additional $k^{3/2}$ results in
\begin{equation} \label{A4_sol2}
k^{3+\eta_4/2}\tilde{A}_{\eta_4}\approx \bar B_{\eta_1} k^{4-\eta_1/2}+\bar{\bar B}_{\eta_1}k^{2-\eta_1/2}\,,
\end{equation}
where we have also assumed $\eta_1>-1$. Squaring this expression we see that the scaling of the power spectrum can now reach $\alpha=8-\eta_1$. In practice, the second term in Eq.~\eqref{A4_sol2} can dominate because these are super-horizon scales, unless the duration of the different stages is tuned. However, this tuning is possible and a power spectrum with a $k^8$ scaling can be found for $\eta_1=\eta_3=0$ and $\eta_2=\eta_4=-6$ as shown in the main text, as well as first found by Ref.~\cite{Tasinato_2021}. For that case, the conditions are
\begin{equation} \label{conditions}
\Delta N_2\lesssim\frac{2}{3}\log\frac{3}{2}\,,\ \  \Delta N_3 > 2\Delta N_2\,,
\end{equation}
where $\Delta N_i$ denotes the duration in e-folds of the stage $i$. Therefore, to make this new scaling appear, one requires a very short first USR stage followed by an SR stage that is at least twice as long. There is a also a requirement that the final USR stage is sufficiently long, so that a sufficient range of scales is affected by both USR stages. In those scenarios, both scalings are typically present, as seen in Fig.~\ref{fig:IF}.

\section*{Acknowledgements}

MWD is supported by a studentship awarded by the Perren Bequest. PC acknowledges support from a UK Research and Innovation Future Leaders Fellowship (MR/S016066/1), and DJM is supported by a Royal Society University Research Fellowship.

\bibliographystyle{JHEPmodplain}
\bibliography{Refs}

\end{document}